\documentclass[12pt]{article}
\usepackage[margin=0.95 in]{geometry}
\usepackage{amsmath} 
\usepackage{slashed}
\usepackage{dsfont}
\usepackage{amssymb,amsfonts}
\usepackage{mathtools}
\usepackage[all]{xy}
\usepackage{graphicx}
\usepackage{amsmath}
\usepackage{amssymb}
\usepackage{float}
\usepackage{array}
\usepackage{tikz}
\usepackage{mathtools}
\usepackage{mathrsfs} 
\usepackage[hidelinks]{hyperref}
\usepackage{cite}
\usepackage{tikz}
\usepackage{array, multirow}  

\numberwithin{equation}{section}
\newcommand{\be}{\begin{equation}}
\newcommand{\ee}{\end{equation}}

\def\bea{\begin{eqnarray}}
\def\eea{\end{eqnarray}}

\numberwithin{equation}{section}
\numberwithin{table}{section}

\begin{document}

\hypersetup{pageanchor=false}
\begin{titlepage}
\vbox{\halign{#\hfil    \cr}}  
\vspace*{15mm}
\begin{center}
{\Large \bf 
Orbifolded Elliptic Genera  of Non-Compact Models
}

\vspace*{10mm} 

{\large 
Sujay K. Ashok$^{a,b}$ and 
Jan Troost$^c$}
\vspace*{8mm}

$^a$The Institute of Mathematical Sciences, \\
		 IV Cross Road, C.I.T. Campus, \\
	 Taramani, Chennai, India 600113

\vspace{.6cm}

$^b$Homi Bhabha National Institute,\\ 
Training School Complex, Anushakti Nagar, \\
Mumbai, India 400094

\vspace{.6cm}

$^c$Laboratoire de Physique de l'\'Ecole Normale Sup\'erieure \\ 
 \hskip -.05cm
 CNRS, ENS, Universit\'e PSL,  Sorbonne Universit\'e, \\
 Universit\'e  Paris Cit\'e  \\
 24 rue Lhomond,
 F-75005 Paris, France	 

\vspace*{0.8cm}
\end{center}

\begin{abstract} { We revisit the flavored  elliptic genus of the N=2 superconformal cigar model and generalize the analysis of the path integral result to the case of real central charge. It gives rise to a  non-holomorphic modular covariant function generalizing completed mock modular forms.  We also compute the genus for  angular orbifolds of the 
cigar and Liouville theory and decompose it in terms of discrete and continuous contributions.  The orbifolded elliptic genus at fractional level is a completed mock modular form with a shadow related to U$(1)$ modular invariants at rational radius squared. We take the limit of the orbifolded genera towards a weighted ground state index and carefully interpret the contributions. We stress that the orbifold cigar and Liouville theories have a maximal and a minimal radius, respectively.
}  
\end{abstract}

\end{titlepage}

\hypersetup{pageanchor=true}

\setcounter{tocdepth}{2}
\tableofcontents

\section{Introduction}
Elliptic genera of two-dimensional conformal field theories have a rich connection to mathematics. For compact theories, they relate to a Dirac loop index of vector bundles over the target space \cite{Schellekens:1986xh,Witten:1986bf}. For non-compact theories, there is a rich connection \cite{Troost:2010ud} to completed mock modular forms \cite{Zwegers}. Through these connections, these genera are relevant to basic features of physical theories such as anomalies in string theory \cite{Schellekens:1986xh} and the microscopic entropy of black holes \cite{Dabholkar:2012nd}. In this paper we wish to further extend this connection by exploring how natural physical questions translate into mathematical expressions that go beyond the standard framework.   

The cigar coset conformal field theory provided the original bridge between non-compact elliptic genera of interacting two-dimensional conformal field theories and completed mock modular forms \cite{Troost:2010ud}. It was analyzed at integer level as a Jacobi form of a modular variable and two angular variables \cite{Ashok:2011cy}, at fractional level for a modular variable and one angular variable \cite{Eguchi:2010cb} as well as orbifolds thereof \cite{Sugawara:2011vg}. In section \ref{EllipticGenus} of this paper, we calculate the elliptic genus for any strictly positive real level as a function of one modular variable and two fugacities. We briefly comment on how this provides a physically motivated generalization of completed mock modular forms. Moreover, we point out that the global U$(1)$ rotation symmetry of the cigar allows for an orbifold by a $\mathbb{Z}_P$ subgroup for any positive integer order $P$, providing a clear interpretation of a generic orbifold \cite{Sugawara:2011vg}. We compute the elliptic genera of the ensuing models in section \ref{Orbifolds} and find that they give rise to completed mock modular forms with a remainder related to a rational U$(1)$ theory. Exploiting and generalizing the analysis of the chiral ring of non-compact Landau-Ginzburg models \cite{Li:2018rcl}, we considerably improve the interpretation of the final result in the case where the level is fractional and carefully distinguish between the state and the operator ring interpretation of the corresponding weighted index in section \ref{WeightedIndex}.  We summarize our results and draw conclusions in section \ref{Conclusions}.  

\section{The  Flavored Elliptic Genus Revisited}
\label{EGcigarNbyK}
\label{EllipticGenus}

We begin our analysis with the path integral expression for the elliptic genus of the cigar coset theory \cite{Troost:2010ud,Eguchi:2010cb,Ashok:2011cy}. The cigar theory has a local ${\cal N}=2$ superconformal symmetry as well as a global U$(1)$ angular rotation symmetry. We study the elliptic genus which includes a fugacity for the angular momentum charge \cite{Ashok:2011cy}. We stress the point that the path integral derivation \cite{Ashok:2011cy} is valid for a generic strictly positive level. We initiate an analysis of the formula and explain why the expression  is a highly non-trivial mathematical object of interest in its own right.

\subsection{The Cigar Elliptic Genus at Generic Level}

A path integral derivation of the cigar elliptic genus at generic level ${k}$  yields the result \cite{Troost:2010ud,Ashok:2011cy}
\begin{eqnarray}
\chi(\tau, \alpha, \beta) =
{k}\, \int_{-
\infty}^{+
\infty} ds_{1} ds_{2}\ 
\frac{\theta_{11} (\tau,s_1\tau +s_2 - \frac{{k}+1}{{k}} \alpha+\beta )}{
\theta_{11}(\tau, s_1 \tau+ s_2 - \frac{\alpha}{{k}} + \beta )}\ 
e^{ - \frac{{k} \pi}{ \tau_2} | s_1  \tau + s_2  |^2} \ 
 e^{-2\pi i\alpha_2(s_1 \tau + s_2)} \, .
\label{pathintegralformula}
\end{eqnarray}
In this expression, the modular parameter of the torus on which we compute the path integral is $\tau$, the fugacity for the R-charge in the ${\cal N}=2$ superconformal algebra is $\alpha$, the fugacity for the cigar angular momentum is $\beta$ and the real numbers $(s_1,s_2)$ code holonomies of the gauge field of the gauged Wess-Zumino-Witten model on the torus, as well as winding numbers. The R-charge fugacity is parameterized by $\alpha=\alpha_1+\tau \alpha_2$, and the angular momentum fugacity is $\beta = \beta_1+ \tau \beta_2$, with $(\alpha_1,\alpha_2)$ and $(\beta_1, \beta_2)$  real numbers. 

The path integral expression for the cigar elliptic genus at fractional level ${k}$ is a neat way to code a completed mock modular form \cite{Troost:2010ud}. In mathematics, alternative expressions for the completed mock modular form exist \cite{Zwegers} -- the path integral result is equally well-defined and renders all of the modular and elliptic properties manifest. Both the mock modular holomorphic part and its completion can be reconstructed from the path integral \cite{Troost:2010ud}.  

When the level ${k}$ is irrational, the cigar elliptic genus (\ref{pathintegralformula}) remains well-defined but the mathematical object it represents falls {\em outside} of the known class of completed mock modular Jacobi forms. Since this is an important point, let us argue how this is seen from a physical perspective. We stress that the modular properties of (\ref{pathintegralformula}) as well as the elliptic properties in the angular charge fugacity $\beta$ remain manifest:
\begin{align}
\chi(\tau+1,\alpha,\beta) &= \chi(\tau,\alpha,\beta)
\nonumber \\
\chi(-\frac{1}{\tau},\frac{\alpha}{\tau},\frac{\beta}{\tau}) &= e^{ \pi i \frac{c}{3} \frac{\alpha^2}{\tau} -2 \pi i \frac{\alpha \beta}{\tau} } \chi(\tau,\alpha,\beta) 
\nonumber \\
\chi(\tau, \alpha,\beta+1) &= \chi(\tau,\alpha,\beta)
\nonumber \\
\chi(\tau,\alpha,\beta+\tau) &= e^{2 \pi i \alpha} 
\chi(\tau,\alpha,\beta)
\, .
\label{ModularAndAngularElliptic}
\end{align}
However, the R-charges of the excitations of the model differ by irrational ratios related to the irrational level ${k}$. This feature of the field theory makes for a lack of ellipticity in the R-charge fugacity $\alpha$ of its elliptic genus. The canonical reasoning for ellipticity is based on a quantization of R-charge which no longer holds at an irrational central charge. These initial remarks are meant to stress that the formula (\ref{pathintegralformula}) encourages further research into a mathematical generalization of completed mock modular forms to include the more general path integral expression (\ref{pathintegralformula}) with the transformation properties (\ref{ModularAndAngularElliptic}) and a subtle  dependence on $\alpha$ that needs to be characterized carefully. We leave this for future work -- see also \cite{Sugawara:2011vg} -- and concentrate on the case where ${k}$ is fractional and we have a path integral completed mock Jacobi form.\footnote{The qualitative difference between the theories with fractional and irrational central charge is ubiquitous in conformal field theory.}

\subsection{The Cigar Elliptic Genus at Fractional Level}
In this subsection we present the transformation properties of the cigar coset elliptic genus at fractional level, including a fugacity for a global symmetry, thereby mildly  extending the results of \cite{Troost:2010ud, Eguchi:2010cb, Ashok:2011cy}. We set the fractional level ${k}$ equal to ${k}=N/D$ where $(N,D)$ are co-prime positive integers representing the numerator and the denominator of the level. In this case, the Jacobian ellipticity properties that complement the transformation rules (\ref{ModularAndAngularElliptic}) are:
\begin{align}
\chi(\tau,\alpha+N,\beta) &= (-1)^{\frac{c}{3}N}
\chi(\tau,\alpha,\beta)
\nonumber \\
\chi(\tau, \alpha+ N \tau, \beta) &=(-1)^{\frac{c}{3}N}
q^{-\frac{c}{6} N^2} e^{-2 \pi i \alpha \frac{c}{3}N}
e^{2 \pi i \beta N}  \chi(\tau,\alpha,\beta)
\, .
\end{align}
We  shift the holonomy variables in equation \eqref{pathintegralformula} by $\beta-\frac{\alpha}{{k}}$ and obtain the path integral formula \cite{Ashok:2014nua}
\begin{eqnarray}
\chi(\tau, \alpha, \beta) =
{k}\, \int_{-
\infty}^{+
\infty} ds_{1} ds_{2}\ 
\frac{\theta_{11} (\tau,s_1\tau +s_2 -\alpha )}{
\theta_{11}(\tau, s_1 \tau+ s_2 )}\ 
e^{ - \frac{{k} \pi}{ \tau_2}(s_1  \tau + s_2 + \frac{\alpha}{{k}} -\beta)(s_1  \bar\tau + s_2 + \frac{\alpha}{{k}} -\bar\beta)} \ 
\, .
\label{pathintegralformulaforexpansion}
\end{eqnarray}
The shift ensures that  the $\theta_{11}$-factor in the denominator is no longer a function of the fugacities. This important detail makes it possible to separate the subtleties of the holonomy integral from the dependence on the fugacities. Thus we can now read off that the dependence of the elliptic genus on the fugacity $\alpha$ is purely holomorphic while there is an anti-holomorphic dependence on the global  charge fugacity $\beta$. This is consistent with the fact that  $\alpha$ is the fugacity for the chiral R-current,  while $\beta$ is the fugacity for the non-chiral angular momentum on the cigar.

\section{Orbifolded Genera}
\label{Orbifolds}

To start off, we stress that the cigar geometry caps off regularly at the tip. Given an asymptotic radius and linear dilaton related to the central charge, there is a single regular filling in of the bulk geometry.\footnote{See \cite{Hori:2001ax} for a renormalization group analysis in this spirit.} A natural generalization  are the geometries in which we allow for an orbifold singularity at the tip. These geometries have a smaller asymptotic radius which is the regular, maximal radius divided by an integer. In turn, the ${\cal N}=2$ Liouville model which has a superpotential that is an exponential of a complex superfield, has a minimal radius. Well-definedness of the exponential potential allows also for a choice of radius of the angular field to be any multiple of the minimal radius. These two facts are T-dual.\footnote{While ``the cigar model" has always referred to the regular geometry in the literature, "the Liouville theory" has often  referred to a model with a non-minimal radius. This detail in nomenclature should be kept in mind when comparing our paper to the literature.} 

Therefore, the cigar at regular as well as Liouville theory at minimal radius are natural starting points for an analysis or classification of physical models and  for a parameterization of completed mock modular forms.
 In this section, to generate old and new models as well as expressions for interesting examples of completed mock modular forms, we orbifold the regular cigar model at fractional level by a discrete $\mathbb{Z}_P$ subgroup of the global angular  U$(1)$. We also refer to  \cite{Sugawara:2011vg} for the one-parameter partition functions of these orbifolds. Including the fugacity for the angular charge clarifies the derivation of the orbifold since the orbifold group is a subgroup of the angular $U(1)$ global symmetry.

\subsection{The Path Integral for Orbifolds of the Coset}

In the path integral formulation of the coset elliptic genus, we can identify the true holonomies of the gauge field to be in the interval $s_i\in [0,1[$ and introduce the integer winding numbers $\tilde m$ and $\tilde w$:
\begin{align}
\chi(\tau, \alpha, \beta) &=
\frac{N}{D}~ \sum_{\tilde m, \tilde w}~ \int_{0}^{1} ds_{1} ds_{2}\ 
\frac{\theta_{11} (\tau,s_1\tau +s_2 -\alpha )}{
\theta_{11}(\tau, s_1 \tau+ s_2 )} ~e^{2\pi i \alpha \tilde w}\nonumber\\
& \hspace{3cm}e^{ - \frac{{k} \pi}{ \tau_2}((s_1+\tilde w)  \tau + (s_2 +\tilde m)+ \frac{\alpha}{{k}} -\beta)(  (s_1+\tilde w)  \bar\tau + (s_2 +\tilde m)+ \frac{\alpha}{{k}} -\bar\beta)} \ 
\, .
\label{pathintegralformulawithmw}
\end{align}
In order to define the genus of the theory orbifolded by the group $\mathbb{Z}_P$, we  introduce $\mathbb{Z}_P$-twisted sectors. In the Lagrangian formulation, since the radius becomes smaller, they are labelled by $\mathbb{Z}_P$ valued integers $(m_a, m_b)$ that parameterize the extra winding sectors along the two directions of the world sheet torus:
\be 
\label{fracwind}
\tilde w = w + \frac{m_a}{P} ~,\qquad \tilde m =  m + \frac{m_b}{P} \, .
\ee 
We sum over the twisted sectors to obtain the orbifolded elliptic genus: 
\begin{align}
\chi^{\mathbb{Z}_P}(\tau, \alpha, \beta)
 &=
\frac{N}{D P }  \sum_{m_a,m_b}
 \int_{0}^1 ds_i \frac{\theta_1(\tau,-s_1 \tau-s_2+\alpha)}{\theta_1(\tau,-s_1 \tau-s_2)}
e^{\frac{2\pi i\alpha (P w + m_a)}{P}} \nonumber\\
&\hspace{1cm} e^{- \frac{N \pi }{ D \tau_2} \big( (s_1 +\frac{P w+m_a}{P})\tau +  s_2 +  \frac{Pm+m_b}{P} + \frac{D\alpha}{N} -\beta \big) \big( (s_1 +\frac{P w+m_a}{P})\bar\tau +  s_2 +  \frac{Pm+m_b}{P} + \frac{D\alpha}{N}-\bar\beta \big) }~.
\end{align}
This is a one-parameter generalization of the partition functions derived in \cite{Sugawara:2011vg}.
A second way to understand the orbifolding procedure is to use the fugacity for the angular momentum generator and to note that we can equivalently write the expression for the orbifolded elliptic genus  as 
\begin{align}
    \chi^{\mathbb{Z}_P}(\tau,\alpha,\beta) = \frac{1}{ P }  \sum_{m_a,m_b =0}^{P-1}~ e^{  \frac{2\pi i\alpha  m_a}{P} }  ~
 \chi(\tau, \alpha, \beta - \frac{m_a \tau+m_b}{P}) ~. 
\end{align}
The integer $m_b$ projects onto $\mathbb{Z}_P$ invariants while the integer $m_a$ is the  twisted sector label. The phase prefactor 
accounts for the fact that the ground state in the $m_a$-twisted sector carries extra $R$-charge since the R-charge has an angular momentum term. 
Instead of explicitly keeping track of the twisted sectors, it is technically simpler to replace $(Pw +m_a, Pm + m_a)$ by a pair of integers, which we once more label $(w,m)$, and write the  elliptic genus of the orbifolded theory as 
\begin{align}
\chi^{\mathbb{Z}_P}(\tau, \alpha, \beta) 
 &=
\frac{N}{D P} 
 \int_{0}^1 ds_i \frac{\theta_1(\tau,-s_1 \tau-s_2+\alpha)}{\theta_1(\tau,-s_1 \tau-s_2)}
e^{\frac{2\pi i \alpha w}{P}} \nonumber\\
&\hspace{3cm} e^{- \frac{N \pi }{ D \tau_2} \big( (s_1 +\frac{w}{P})\tau +  s_2 +  \frac{m}{P} + \frac{D\alpha}{N} -\beta \big) \big( (s_1 +\frac{w}{P})\bar\tau +  s_2 +  \frac{m}{P} + \frac{D\alpha}{N}-\bar\beta \big) } \, .
\end{align}

\subsubsection{Modular Properties of the Orbifolded Genus}
We  check the modular properties of the orbifolded elliptic genus. To make this easier, we rewrite the genus as:
\begin{align}
\chi^{\mathbb{Z}_P}(\tau, \alpha, \beta) 
&= \frac{N}{D P } 
 \int_{0}^1 ds_i \frac{\theta_1(\tau,-s_1 \tau-s_2+\alpha)}{\theta_1(\tau,-s_1 \tau-s_2)}
e^{\frac{2\pi i \alpha w}{P}}  \nonumber\\
& \hspace{3cm}e^{- \frac{N \pi }{ D \tau_2} \left|  (s_1 +\frac{w}{P})\tau +  s_2 +  \frac{m}{P} + \frac{D\alpha}{N} -\beta\right|^2 }e^{-2 \pi i \alpha_2 ((s_1 + \frac{w}{P})\tau +s_2+ \frac{m}{P} + \frac{D\alpha}{N} -\beta ) } ~.
\label{chiorbfinalline}
\end{align}
Under the $T$-transformation, we have
\begin{align}
   T:\quad \tau&\rightarrow \tau + 1~, \quad \alpha_1 \rightarrow \alpha_1 - \alpha_2~, \quad s_1 \rightarrow s_1 - s_2~,\quad 
    m \rightarrow m-w~ \, ,
\end{align}
and  the elliptic genus remains invariant. Under the S-transformation, we have
\begin{align}
   S:\quad \tau &\rightarrow -\frac{1}{\tau}~, \quad \alpha\rightarrow \frac{\alpha}{\tau}~,\quad s_1\rightarrow -s_2~, \quad s_2 \rightarrow s_1~,\quad m\rightarrow w~,\quad w\rightarrow m ~.
\end{align}
The transformation on the $\alpha=\alpha_1+\alpha_2 \tau$ parameter is equivalent to the map
\be 
\alpha_1 \rightarrow \alpha_2~, \quad \alpha_2\rightarrow  \alpha_2\tau - \alpha~.
\ee 
Under the S-transformation, the absolute value squared in the exponent of equation \eqref{chiorbfinalline} remains invariant. The remaining terms picks up the phase factor
\begin{equation}
    e^{\frac{\pi i}{\tau}(\alpha^2 - 2\alpha(s_1\tau+s_2))}~e^{-2\pi i \frac{\alpha}{P\tau} (w\tau+  m)}~ e^{2\pi i \frac{\alpha}{\tau}((s_1 + \frac{w}{P})\tau + \frac{m}{P}+s_2 + \frac{D\alpha}{N}-\beta)}
    = e^{ \frac{\pi i \alpha^2}{\tau}(1 + \frac{2D}{N})}~e^{-\frac{2\pi i \alpha\beta}{\tau}}~.
\end{equation}
Thus, we can summarize the modular transformation properties
\begin{align}
\chi^{\mathbb{Z}_P}(\tau+1, \alpha, \beta) &= \chi^{\mathbb{Z}_P}(\tau, \alpha, \beta)~,\\
\chi^{\mathbb{Z}_P}(-\frac{1}{\tau}, \frac{\alpha}{\tau}, \frac{\beta}{\tau}) &= e^{ \frac{\pi i c }{3}\frac{\alpha^2}{\tau}} ~e^{-\frac{2\pi i \alpha\beta}{\tau}}~ \chi^{\mathbb{Z}_P}(\tau, \alpha, \beta)~.
\end{align} 
This is identical to the modular transformation of the cigar elliptic genus with central charge $c=3+6\frac{D}{N}$. 

\subsubsection{Elliptic Properties of the Orbifolded Genus}

We move on to determine the ellipticity properties of the orbifolded genus.
Firstly, we check under which shifts  $L$
\be
\alpha \rightarrow \alpha+ L~
\ee
of the R-charge fugacity $\alpha$ the genus is covariant.
We propose to compensate the shift by a change of summation variable:
\be
 m \rightarrow m - \frac{D P}{N} L~.
\ee
The combined changes ensure that the argument of the absolute value squared term in the exponent of (\ref{chiorbfinalline}) remains invariant. In order for the shift of the integer $m$ to be integer as it must, and given that the greatest common divisor of $N$ and $D$ is one, we conclude that $N$ has to be a divisor of $LP$. The remaining factors in \eqref{chiorbfinalline} pick up a phase equal to
$
(-1)^L e^{\frac{2\pi i L w}{P} }
$, 
and for this phase to be independent of the winding number, we  require that the order $P$ has to be a divisor of the shift $L$. 
If both these conditions 
\be
N| LP \,  \qquad \text{and} \qquad P|L
\label{ConditionsOnShift}
\ee
are satisfied, then we have that the orbifolded elliptic genus is  multiplied by a factor $(-1)^L$ under the $\alpha$-shift by $L$. 
Next we consider the combined shift 
\be
\alpha \rightarrow \alpha+ \tilde{L} \tau~, \qquad \qquad w \rightarrow w - \frac{D \tilde{L}}{N} P~.
\ee
We find that the integer $N$ has to be a divisor of $\tilde{L} P$.  
The remaining terms in the expression \eqref{chiorbfinalline} pick up the factor
\begin{align}
&
(-1)^{\tilde L} q^{-\frac{\tilde L^2}{2}(1+\frac{2D}{N})} ~ e^{-2\pi i \alpha \tilde L (1+ \frac{2D}{N})} ~e^{2\pi i\beta \tilde L }~ e^{-\frac{2\pi i m \tilde L}{P}}~.
\end{align}
In order for the phase factors to be independent of the winding numbers, we require the order $P$ to be a divisor of the shift $\tilde{L}$. Thus, the conditions on the variable $\tilde{L}$ are identical to those on the integer $L$ found previously. We summarize the ellipticity properties:
\begin{align}
    \chi^{\mathbb{Z}_P}(\tau, \alpha + L, \beta) &=  (-1)^L \chi^{\mathbb{Z}_P}(\tau, \alpha, \beta) \\ 
     \chi^{\mathbb{Z}_P}(\tau, \alpha + L\tau, \beta) &= (-1)^L\, q^{-\frac{c L^2}{6}}\, e^{-\frac{2\pi i c \alpha L}{3}}\, e^{2\pi i \beta L}\, \chi^{\mathbb{Z}_P}(\tau, \alpha, \beta)~
\end{align}
under the conditions (\ref{ConditionsOnShift}).

Lastly, we turn to the ellipticity in the $\beta$-variable. From the expression \eqref{chiorbfinalline} it is clear that a shift of the angular momentum fugacity $\beta$ by $\frac{1}{P}$ can be compensated by a shift of the $m$ winding number. Similarly a shift of $\beta$ by $\frac{\tau}{P}$ can be compensated by a shift of the winding $w$. In the latter case, this leads to an overall $\alpha$-dependent phase. We therefore have the ellipticity: 
\begin{align}
    \chi^{\mathbb{Z}_P}(\tau, \alpha,\beta+\frac{1}{P}) &= \chi^{\mathbb{Z}_P}(\tau,\alpha,\beta)
\\
\chi^{\mathbb{Z}_P}(\tau,\alpha,\beta+\frac{\tau}{P}) &= e^{\frac{2 \pi i \alpha}{P}} 
\chi^{\mathbb{Z}_P}(\tau,\alpha,\beta)
\, .
\end{align}
The angular orbifold has $\mathbb{Z}_P$ projected the angular momenta such that the periodicity in the corresponding fugacity is $P$ times smaller. The R-symmetry fugacity periodicity is a function of the quantization of the left R-charges which is a considerably more subtle function of the order of the diagonal orbifold, as described above. 
This concludes the analysis of the modular and elliptic properties of the orbifolded genus. 

\subsection{The Spectrum}

In this section we rewrite the orbifolded elliptic genus in its Hamiltonian form to identify the contributions of different parts of the spectrum of the theory. We identify a holomorphic contribution and a non-holomorphic remainder term. The holomorphic part arises from a sum over extended discrete characters and therefore from a tower of states built on discrete states and their spectral flows. The non-holomorphic part arises from the continuous part of the spectrum through a mismatch of bosonic and fermionic spectral densities \cite{Troost:2010ud,Ashok:2011cy}. Our orbifold partition function displays an interesting interplay between the integers $(N,D)$ appearing in the rational radius squared and the order $P$ of the orbifold. In other aspects, the calculation follows the pattern in \cite{Troost:2010ud,Eguchi:2010cb,Ashok:2011cy}.
We recall the orbifolded elliptic genus in the Lagrangian form, in which we have a sum over windings $(m,w)$:
\begin{align}
\chi^{\mathbb{Z}_P}(\tau, \alpha, \beta )&=
\frac{N}{DP} 
\sum_{m,w}
 \int_{0}^1 ds_i \frac{\theta_1(\tau,-s_1 \tau-s_2+\alpha)}{\theta_1(\tau,-s_1 \tau-s_2)}
e^{2\pi i \alpha\frac{w}{P}} \nonumber\\
&\hspace{2.5cm} e^{- \frac{N \pi }{ D \tau_2} \big( (s_1 +\frac{w}{P})\tau +  s_2 +  \frac{m}{P} + \frac{D\alpha}{N} -\beta\big) \big( (s_1 +\frac{w}{P})\bar\tau +  s_2 +  \frac{m}{P} + \frac{D\alpha}{N} -\bar\beta\big) }~.
\end{align}
To go over to the Hamiltonian formulation, we Poisson resum in the $m$ variable to obtain 
\begin{align}
\chi^{\mathbb{Z}_P}(\tau, \alpha, \beta )&=
\sqrt{\frac{N\tau_2}{D}}
\sum_{n,w}
 \int_{0}^1 ds_i \frac{\theta_1(\tau,-s_1 \tau-s_2+\alpha)}{\theta_1(\tau,-s_1 \tau-s_2)}
 e^{2\pi i P n s_2} ~ q^{n(Ps_1+w)}
\nonumber\\
&\hspace{2cm} 
e^{2\pi i \alpha(\frac{w}{P} + \frac{D}{N} nP)} ~ e^{-2\pi i \beta P n}  ~(q\bar q)^{-\frac{DP^2}{4N}\big(n-\frac{N}{DP}(\frac{w}{P} + s_1-\beta_2)\big)^2}~.
\end{align}
The quantum numbers $(n,w)$ play the role of momentum and winding around the asymptotic circle direction.  
We  use the series expansion that is valid for $0 < s_1 < 1$
\begin{equation}
\frac{\theta_1(\tau,-s_1 \tau-s_2+\alpha)}{\theta_1(\tau,-s_1 \tau-s_2)} = -\frac{i \theta_1(\tau,\alpha)}{\eta(\tau)^3} 
\sum_{\ell \in \mathbb{Z}} \frac{z q^{\ell}}{1-zq^{\ell}}
e^{-2 \pi i (s_1 \tau_1+s_2) \ell + 2 \pi s_1 \tau_2 \ell} ~, 
\end{equation}
where $z=e^{2 \pi i \alpha}$
and rewrite the elliptic genus:
\begin{align}
\chi^{\mathbb{Z}_P}(\tau, \alpha, \beta )&=
-\frac{i \theta_1(\tau,\alpha)}{\eta(\tau)^3}
\sqrt{\frac{N\tau_2}{D}}
\sum_{n,w}
 \int_{0}^1 ds_i 
 \sum_{\ell \in \mathbb{Z}} \frac{z q^{\ell}}{1-zq^{\ell}}
e^{-2 \pi i (s_1 \tau_1+s_2) \ell + 2 \pi s_1 \tau_2 \ell}
e^{2\pi i P n s_2} \nonumber\\
&\hspace{1cm} q^{n(Ps_1+w)}~ e^{2\pi i \alpha(\frac{w}{P} + \frac{D}{N} nP)} ~ e^{-2\pi i \beta P n}
~(q\bar q)^{-\frac{DP^2}{4N}\big(n-\frac{N}{DP}(\frac{w}{P} + s_1-\beta_2)\big)^2}~.
\end{align}
The integral over the holonomy $s_2$ glues the variable $\ell$ to the momentum $n$:
\be 
\ell = Pn~ .
\ee  
Substituting for the variable $\ell$ we get
\begin{align}
\chi^{\mathbb{Z}_P}(\tau, \alpha, \beta )&=
-\frac{i \theta_1(\tau,\alpha)}{\eta(\tau)^3}
\sqrt{\frac{N\tau_2}{D}}
\sum_{n,w}
 \frac{z q^{Pn}}{1-zq^{Pn}}
 \int_{0}^1 ds_1 e^{-\frac{N}{D}\pi \tau_2 s_1^2 +\frac{2\pi s_1\tau_2}{D}(D P n -N\frac{w}{P}+ N\beta_2))}
 \nonumber\\
&\hspace{1cm} q^{nw}~ e^{\frac{2\pi i \alpha}{N}( DP n + N\frac{w}{P} )} ~ e^{-2\pi i \beta P n}
~(q\bar q)^{-\frac{1}{4ND}\big(DPn-\frac{N}{P}w + N\beta_2)\big)^2}~.
\end{align}
We  introduce the integral over radial momentum $s$: 
\begin{align}
e^{-\pi \tau_2 \frac{N}{D} s_1^2} = \sqrt{\frac{\tau_2}{ND}}\int_{\mathbb{R} - i\epsilon} ds ~e^{-\frac{\pi}{ND} \tau_2 s^2 - 2\pi i \tau_2 \frac{s_1}{D} s} 
\end{align}
and obtain 
\begin{align}
\chi^{\mathbb{Z}_P}(\tau, \alpha, \beta )&=
-\frac{i \theta_1(\tau,\alpha)}{\eta(\tau)^3}
\frac{\tau_2}{D}
\sum_{n,w}
 \frac{z q^{Pn}}{1-zq^{Pn}}
 \int_{\mathbb{R} - i\epsilon} ds ~e^{-\frac{\pi}{ND} \tau_2\Big (s^2+\big(DPn-\frac{N}{P}w + N\beta_2)\big)^2\Big) } \nonumber \\
&\hspace{1cm} \int_{0}^1 ds_1  e^{- 2\pi i \tau_2 \frac{s_1}{D} (s+i(D P n -N\frac{w}{P}+ N\beta_2))}
 ~ q^{nw}~ e^{\frac{2\pi i \alpha}{N}( DP n + N\frac{w}{P} )} ~ e^{-2\pi i \beta P n}~.
\end{align}
We perform the $s_1$ integral
and find:
\begin{align}
\chi^{\mathbb{Z}_P}(\tau, \alpha, \beta )&=
-\frac{1}{2 \pi }
\frac{ \theta_1(\tau,\alpha)}{\eta(\tau)^3}
\sum_{n,w}
 \frac{z q^{Pn}}{1-zq^{Pn}}
 \int_{\mathbb{R} - i\epsilon} ds ~e^{-\frac{\pi}{ND} \tau_2\Big (s^2+\big(DPn-\frac{N}{P}w + N\beta_2)\big)^2\Big) } \nonumber \\
&\hspace{1cm} 
\frac{1- e^{- 2\pi i \tau_2 \frac{1}{D} (s+i(D P n -N\frac{w}{P}+ N\beta_2))}}{s+i (DPn-N \frac{w}{P}+N \beta_2)}
 ~ q^{nw}~ e^{\frac{2\pi i \alpha}{N}( DP n + N\frac{w}{P} )} ~ e^{-2\pi i \beta P n}~.
\end{align}
We rewrite the second term in terms of the complexified momentum $\tilde{s}$:
\begin{equation}
\tilde{s} = s + i N
\end{equation}
and simultaneously shift $w \rightarrow w-P$. The shifts leave the denominator invariant. The exponent picks up extra factors such that the elliptic genus can be written as the difference of shifted contours -- we denote $\tilde{s}$ again as $s$ --:
\begin{align}
\chi^{\mathbb{Z}_P}(\tau, \alpha, \beta )&=
-\frac{\theta_1(\tau,\alpha)}{2 \pi\eta(\tau)^3}
\sum_{n,w}
 \frac{1}{1-zq^{Pn}}
 \left(
 \int_{\mathbb{R} - i\epsilon} ds~z q^{Pn} - \int_{\mathbb{R}+iN - i\epsilon} ds
 \right)  \nonumber \\
&\hspace{1cm} 
\times\frac{e^{-\frac{\pi}{ND} \tau_2\Big (s^2+\big(DPn-\frac{N}{P}w + N\beta_2)\big)^2\Big) }}{s+i (DPn-N \frac{w}{P}+N \beta_2)}
 ~ q^{nw}~ e^{\frac{2\pi i \alpha}{N}( DP n + N\frac{w}{P} )} ~ e^{-2\pi i \beta P n}~.
 \label{SumOfContours}
\end{align}
The dependence of the pole contributions is on $N\beta_2-\epsilon$. The choice of picking a small positive $\epsilon$ is equivalent to choosing a small negative $N\beta_2$. In the subsequent analysis we shall set $\beta_2=0$ and choose a small positive $\epsilon$. The flavored genera with a different value of the complexified angular fugacity $\beta_2$ are related to our elliptic genus via a wall crossing phenomenon discussed in detail in \cite{Ashok:2014nua}.

\subsubsection{The Discrete Part of the Spectrum}
We shift the second contour in equation (\ref{SumOfContours}) by $iN$ to bring it back to the real axis and identify the resulting contributions as arising from the real momenta of a continuous spectrum. In the process, we pick up the poles that make up the discrete spectrum contribution to the elliptic genus. The poles lie at values of $-is$ equal to the right  momenta $DPn-Nw/P$. They lie between $-i \epsilon$  and $i(N-\epsilon)$. 

To characterize and enumerate the poles, we make a simplifying assumption from here onwards. We assume that the greatest common divisor of the order $P$ and the level numerator $N$ is equal to one. As a consequence we have that  $\text{g.c.d}(DP^2,N)=1$. Therefore, we pick up $NP$ poles. We can write the result in terms of the integer $v$ proportional to the right-moving circle momentum
\begin{equation}
v  =Nw - DP^2 n\, .
\end{equation}
Let us denote a fixed solution of the Diophantine equation
$1=Nw-DP^2n$ as $(n_0,w_0)$. We can pick $n_0$ between $0$ and $N-1$ to make it unique. The solutions for $v$ are then:
\be
(n,w) \in v(n_0,w_0)+(N,DP^2) \mathbb{Z} \, .
\ee 
We first substitute $Nw$ as a function of $v$ in the pole contribution:
\begin{align}
\chi_{hol}^{\mathbb{Z}_P}(\tau, \alpha, \beta )&=-
\frac{i\theta_1(\tau,\alpha)}{\eta(\tau)^3}
\sum_{v=0}^{NP-1}\sum_{n}'
 \frac{1}{1-zq^{Pn}}
 ~ q^{\frac{n}{N} (DP^2 n + v)}~ e^{\frac{2\pi i \alpha}{N}( 2DP n  + \frac{v}{P} )} ~ e^{-2\pi i \beta P n}~.
\end{align}
The prime on the $n$ summation indicates that we have $n=v n_0 + N \ell$ where $\ell$ parameterizes the solutions of the Diophantine equation. This implies:
\begin{align}
\chi_{hol}^{\mathbb{Z}_P}(\tau, \alpha, \beta )
 &=-
\frac{i \theta_1(\tau,\alpha)}{\eta(\tau)^3}
\sum_{v=0}^{NP-1}\sum_{\ell \in \mathbb{Z}}
 \frac{ (z\, q^{P(v n_0+ N\ell)})^{\frac{v}{NP} }}{1-zq^{P(v n_0+N \ell)}}
 ~ q^{N D P^2 (\frac{v n_0}{N}+ \ell)^2}~z^{2DP (\frac{ v n_0}{N}+\ell)  } ~ y^{- P (v n_0+N \ell)}~,
\end{align}
where $y=e^{2 \pi i \beta}$.
From the definitions in Appendix \ref{extendedcharacters}, it is clear that what we can write the discrete contribution to the elliptic genus as  a sum over flavored and extended discrete characters of the ${\cal N}=2$ superconformal algebra: 
\begin{align}
\chi_{hol}^{\mathbb{Z}_P}(\tau,\alpha,\beta) &=  
\sum_{v=0}^{NP-1} \text{Ch}^{(\tilde R)}_{d}(NP, v,Pvn_0;\tau,\alpha, \beta) \, .
\end{align}
The holomorphic part of the elliptic genus receives contributions from Ramond-Ramond ground states and left-moving descendants and their spectral flows with an R-charge fixed by the common imaginary radial momentum. We sum over all R-charges between $-1/2$ and $+1/2$. The spectral flow intercept $Pvn_0$ is a related to the charge through a Diophantine dependence determined by a solution to the equation $v=N w_0 -DP^2 n_0$.

\subsubsection{The Continuous Part of the Spectrum}

We  return to equation \eqref{SumOfContours} and read off the non-holomorphic remainder part. It is given by the line integral: 
\begin{align}
\chi_{rem}^{\mathbb{Z}_P}(\tau, \alpha, \beta )&=
-\frac{1}{2 \pi  }
\frac{ \theta_1(\tau,\alpha)}{\eta(\tau)^3}
\sum_{n,w}
 \frac{1}{1-zq^{Pn}}
 \int_{\mathbb{R} - i\epsilon} ds~(z q^{Pn} -1)   \nonumber \\
&\hspace{3cm} 
\frac{e^{-\frac{\pi}{ND} \tau_2\Big (s^2+\big(DPn-\frac{N}{P}w )\big)^2\Big) }}{s+i (DPn-N \frac{w}{P})}
 ~ q^{nw}~ e^{\frac{2\pi i \alpha}{N}( DP n + N\frac{w}{P} )} ~ e^{-2\pi i \beta P n}
 \nonumber \\
&=  
\frac{ \theta_1(\tau,\alpha)}{2\pi\eta(\tau)^3}
\sum_{m,\bar{m}}
 \int_{\mathbb{R} - i\epsilon} \frac{ds}{s+i (DPn-N \frac{w}{P})}~ e^{\frac{2\pi i \alpha}{N}( DP n + N\frac{w}{P} )} ~ e^{-2\pi i \beta P n}  \nonumber \\
&\hspace{5cm} 
\times q^{\frac{s^2}{4 ND}+\frac{(DPn+Nw/P)^2}{4 ND}} \bar{q}^{ \frac{s^2}{4 ND}+\frac{(DPn-Nw/P)^2}{4 ND}}
 ~.
\end{align}
In the exponent of the nomes $q$ and $\bar q$, we recognize the contribution due to the radial momentum as well as the momentum modes  on a circle of radius\footnote{If $N$ and $P^2$ have a common factor $G$, the orbifold radius in equation (\ref{OrbifoldRadius}) can be simplified. This will lead subsequently to a description in terms of level $N D P^2/G^2$ theta functions. We stick to the case $G=1$ for simplicity of presentation.}
\be 
R_{orb} = \sqrt{\frac{\alpha' N}{P^2D}}~. \label{OrbifoldRadius}
\ee 
Our goal now  is to show that one can write the contribution from the continuous part of the spectrum in terms of the same left/right spectral combinations that appear in the partition function of a compact boson at radius  $R_{orb}$. See for instance \cite{DiFrancesco:1987gwq,Cappelli:1986hf,Gepner:1986hr} for background. To do so, we introduce the substitutions \cite{DiFrancesco:1997nk}:
\be\begin{aligned}
DP^2 n + N w &= 
2 N DP^2 u + m \\
DP^2 n - N w &= 
2 N DP^2 \bar{u} + \bar{m} \, ,
\end{aligned}
\ee
with $u,\bar{u}$ integers and $m,\bar{m} \in \mathbb{Z}_{2 N DP^2}$
and find:
\begin{multline}
\chi_{rem}^{\mathbb{Z}_P}(\tau,\alpha,\beta) 
= 
\frac{\theta_1(\tau,\alpha)}{ 2 \pi \eta(\tau)^3}
\sum_{\substack{m-\bar{m}\, \in \,  2N\mathbb{Z}\\ \, \, \, m+\bar{m}\, \in\,  2DP^2\mathbb{Z}}}
\sum_{u,\bar{u}}
 \int_{\mathbb{R} - i\epsilon} \frac{ds}{s+ \frac{i}{P} (2N DP^2 \bar{u} + \bar{m})}  (q\bar{q})^{ \frac{s^2}{4 ND}} 
 \\
 e^{\frac{2\pi i \alpha}{N P}(2 N DP^2 u + m )} ~ e^{-2\pi i \frac{\beta}{2DP}  ( 2DN P^2 (u+ \bar{u}) + (m+\bar{m}))}
q^{P^2 ND (u+\frac{m}{2 P^2 DN})^2} \bar{q}^{P^2 ND (\bar{u}+\frac{\bar{m}}{2 P^2 DN})^2} ~.
\end{multline}
We use the definition of a finite-index $\Theta$-function
\be
\Theta_{m,k}(\tau, \alpha) = \sum_{u\in \mathbb{Z}} q^{k(u+\frac{m}{2k})^2}~e^{2\pi i k\alpha (u+\frac{m}{2k})}~,
\ee 
to rewrite the sum over the left-moving variable $u$, and we obtain:
\begin{align}
\chi_{rem}^{\mathbb{Z}_P}(\tau, \alpha, \beta ) 
 =&  
\frac{ \theta_1(\tau,\alpha)}{2\pi\eta(\tau)^3}
\sum_{\bar{u}}\sum_{\substack{m-\bar{m}\, \in \,  2N\mathbb{Z}\\ \, \, \, m+\bar{m}\, \in\,  2DP^2\mathbb{Z}}}
 \int_{\mathbb{R} - i\epsilon} \frac{ds}{s+ \frac{i}{P} (2N DP^2 \bar{u} + \bar{m})} (q\bar{q})^{ \frac{s^2}{4 ND}} \nonumber \\
 &\hspace{1cm}
\bar{q}^{P^2 ND (\bar{u}+\frac{\bar{m}}{2 P^2 DN})^2}
  ~ e^{-2\pi i \beta \frac{1}{2DP} (2 N DP^2  \bar{u} +\bar{m})}
\Theta_{m, NDP^2}(\tau,\frac{ 2\alpha}{NP}-\frac{ \beta}{DP}) \nonumber\\
&=  
\frac{i\theta_1(\tau,\alpha)}{2\eta(\tau)^3}
\sum_{\bar{u}}
\sum_{\substack{m-\bar{m}\, \in \,  2N\mathbb{Z}\\ \, \, \, m+\bar{m}\, \in\,  2DP^2\mathbb{Z}}}
\Theta_{m, NDP^2}(\tau,\frac{ 2\alpha}{NP}-\frac{ \beta}{DP})
~e^{-2\pi i \beta \frac{1}{2DP} (2 N DP^2  \bar{u} +\bar{m})}\nonumber\\
&\times\frac{1}{\pi i }\int_{\mathbb{R} - i\epsilon} \frac{ds}{s+ i (2N DP^2 \bar{u} + \bar{m})}  e^{- \frac{\pi \tau_2}{NDP^2} (s^2+(2NDP^2\bar{u}+\bar{m})^2 )}
{q}^{-\frac{(2NDP^2\bar{u}+\bar{m})^2}{4NDP^2}} ~.
\end{align}
We define the non-holomorphic function \cite{Zwegers}
\begin{equation}
R_{\bar{m}, \kappa}(\tau, \zeta) =\frac{1}{\pi i}\sum_{\lambda\in \bar m + 2 \kappa \mathbb{Z}} e^{2\pi i \zeta \lambda}
\int_{\mathbb{R}-i\epsilon}~ds~ \frac{e^{-\frac{\pi\tau_2}{\kappa}(s^2 + \lambda^2)}}{s- i \lambda}~q^{-\frac{\lambda^2}{4\kappa}} ~,
\end{equation}
 in order to write the remainder in the compact form:
\begin{align}
\chi_{rem}^{\mathbb{Z}_P}(\tau, \alpha, \beta )&=  \frac{1}{2 } 
\frac{i \theta_1(\tau,\alpha)}{\eta(\tau)^3}
\sum_{\substack{m-\bar{m}\, \in \,  2N\mathbb{Z}\\ \, \, \, m+\bar{m}\, \in\,  2DP^2\mathbb{Z}}}
\Theta_{m,NDP^2}(\tau,\frac{ 2\alpha}{NP}-\frac{ \beta}{DP}) R_{\bar m, NDP^2}(\tau, \frac{\beta}{2DP})~.
 \end{align}
We see that the remainder term  has a non-diagonal left-right spectral combination that also appears in the partition function of the rational compact boson conformal field theory at infinity. This is a generalization of the diagonal combination that has appeared in the mathematics and physics literature\footnote{In the context of orbifolds of the fractional level cigar, the only case that is discussed in the literature \cite{Eguchi:2010cb} is the case when $P=N$. Then, the square of the radius is  $\frac{\alpha'}{ND}$ leading to a diagonal combination of left and right sectors.} \cite{Zwegers,Troost:2010ud,Eguchi:2010cb,Ashok:2011cy}.

\subsection{A Perspective on the Angular Orbifold}

The global symmetry orbifold has introduced twisted sectors which are associated to new winding numbers. In the compact ${\cal N}=2$ minimal model, the twisted sectors of the global symmetry orbifolds  can be found by the technique of shifting the R-charge fugacity \cite{Kawai:1993jk} because the twisted sector representations are again part of the standard spectrum of ${\cal N}=2$ minimal models. One can think of this as due to the limited set of choices for unitary representations of minimal models.  In the non-compact orbifolded model, the twisted sectors  lie outside the spectrum of the original model. Thus the technique to find the twisted sectors that we employed is different: we had to introduce fractional windings in the Lagrangian path integral description. If one has a $\mathbb{Z}_P$ orbifolded cigar partition function at hand, then the $\mathbb{Z}_M$ orbifold with $M|P$ can again be obtained by shifting R-symmetry fugacities  \cite{Kawai:1993jk}. An example is the non-minimal Liouville partitions function at central charge $c=3+6D/N$ which is obtained by orbifolding by $\mathbb{Z}_N$, as discussed in \cite{Eguchi:2010cb, Ashok:2011cy}. Once one has this partition function, one can orbifold it by $\mathbb{Z}_M$ to obtain the partition function orbifolded by $\mathbb{Z}_{P=M/N}$. These form a very specific subclass of orbifolded elliptic genera. All of these partitions functions  are therefore greatly generalized by our equally elementary technique of explicitly parameterizing the necessarily new twisted sectors and representations of the non-compact ${\cal N}=2$ superconformal algebra. 

We have stressed a perspective in which we have a  cigar conformal field theory at rational radius squared regularly capping off or an angular orbifold thereof. The regular model has a maximal radius and all other models have a smaller radius. The regular model is the natural starting point for the analysis.\footnote{The  Appell-Lerch sum ${\cal A}_{2k}$ as defined in mathematics corresponds to the elliptic genus of a $\mathbb{Z}_k$ orbifold of a regular cigar model at radius squared equal to the integer $k$.} All other models are obtained by integer order orbifolds and they require the explicit introduction of superconformal representations that lie outside the spectrum of the original model. 
The general treatment shows that there is a class of completed mock modular forms which is related to the set of modular invariant compact boson models at rational radius squared.

\section{Weighted Indices}
\label{WeightedIndex}
In the quest to  interpret the elliptic genus, it is useful to take the limit $\tau_2 \rightarrow \infty$ in order to identify the Ramond sector ground states that contribute, weighted by their R-charges \cite{Lerche:1989uy}. It turns out that the easiest way to calculate the limit is on the initial path integral expression rather than its spectral decomposition. This is due to the mixing between discrete and continuous contributions and the ambiguity in the split between holomorphic and non-holomorphic parts. These subtleties make for an end result that is interesting in its own right.

\subsection{The Cigar Coset Weighted Index}
We calculate the weighted index for the cigar coset at arbitrary level. 
Our starting point is the equation:
\begin{align}
   \chi(\tau,\alpha,\beta) =&\sqrt{{k} \tau_2}\, \int_{0}^{1} ds_{1} ds_{2}\ \sum_{n,w}
\frac{\theta_{11} (\tau,s_1\tau +s_2- \alpha)}{
\theta_{11}(\tau, s_1 \tau+ s_2)} 
e^{2\pi i \alpha(w+\frac{n}{{k}})} e^{-2\pi i n \beta}\nonumber\\
&\hspace{4cm}e^{2\pi i n s_2} \, q^{n (s_1+w)} (q\bar q)^{\frac{1}{4{k}}(n - {k} (s_1+w -\beta_2) )^2} ~. 
\end{align}
In the $\tau_2 \rightarrow \infty $ limit, we can concentrate on the $n=0$ term and we  consider the $s_1$ integral near the point $s_1+w-\beta_2=0$ for each winding $w$ and given small $\beta_2$. 
We distinguish three cases. Firstly, we suppose that $\beta_2=0$. It is then  more convenient to rewrite the integral in terms of the parameter $r=s_1+w$.  There are two contributions near $r=0$: one that comes from the $s_1\rightarrow 0$ limit, in which the $w=0$ state contributes, and another arises from the $s_1\rightarrow 1$ limit, in which the $w=-1$ state contributes. We find:
\begin{align}
I_{\text{cigar}} = \lim_{\tau_2 \rightarrow \infty} \chi &= \lim_{\tau_2 \rightarrow \infty}
\sqrt{{k} \tau_2} \int_0^1 ds_2 \int_{-\epsilon}^{+\epsilon} dr 
\frac{\sin \pi (s_2+r \tau -\alpha )}{\sin \pi (s_2 + r \tau)}
e^{-{k} \pi \tau_2 r^2} \, .
\end{align}
To take the limit further, we need to distinguish the behaviour of the integrand according to the sign of $r$.
 We write:
 \begin{align}
I_{\text{cigar}} &= \lim_{\tau_2 \rightarrow \infty}
\sqrt{{k} \tau_2} \int_0^1 ds_2 \int_{0}^{+\epsilon} dr 
\left(\frac{\sin \pi (s_2+r \tau -\alpha )}{\sin \pi (s_2 + r \tau )}+
\frac{\sin \pi (s_2-r \tau -\alpha)}{\sin \pi (s_2 - r \tau )}\right)
e^{- {k} \pi \tau_2 r^2} 
\nonumber \\
&= \lim_{\tau_2 \rightarrow \infty}
2 \sqrt{{k} \tau_2} \int_0^1 ds_2 \int_{0}^{+\epsilon} dr 
 \cos \pi \alpha
e^{- {k} \pi \tau_2 r^2} 
=\lim_{\tau_2 \rightarrow \infty} \text{Erf} (\epsilon \sqrt{{k}\pi \tau_2}) \cos \pi \alpha
\nonumber \\
&= \cos \pi \alpha
= \frac{1}{2} e^{-\pi i \alpha} + \frac{1}{2} e^{ i \pi \alpha}
\, .
 \end{align}
 We set a fixed cut-off $\epsilon$ on the integration region of $r$ that can contribute, compute the integral and then take the limit $\tau_2 \rightarrow \infty$. The end result is manifestly independent of the finite but small regulator $\epsilon$. The interpretation of the end result is subtle. 
 The left-moving Ramond ground state at charge $- 1/2$  contributes with a multiplicity of one half to the elliptic genus as does the Ramond ground state at charge $+1/2$. That is possible because the model is non-compact and has a non-compact spectrum \cite{Troost:2010ud}. Both these ground states touch the continuum. The (unweighted) supersymmetric index equals one -- this masks the subtle nature of the weighted answer.  We note that the end result preserves the $\alpha \leftrightarrow - \alpha$ R-charge conjugation symmetry. The infinite level interpretation of the result was  discussed in  \cite{Giveon:2014hfa} and the weighted index is independent of the level.

For the cases $\beta_2 \neq 0$ similar reasonings lead to a weighted index equal to 
\be
I_{\text{cigar}} =\begin{cases}
e^{-\pi i \alpha}~\quad&\text{for $\beta_2 < 0$} \\
e^{i\pi \alpha}~\quad&\text{for $\beta_2 > 0$} \, ,
\end{cases}
\ee
which respectively correspond to the contribution from the winding $w=-1$ and $w=0$ states. For further discussion of the dependence of the spectrum of discrete states on the complex part of the angular fugacity, see \cite{Ashok:2014nua}.

 We add a few further comments. It is important to realize that the elliptic genus is non-zero because of a cancellation in the right-moving sector between a fermion zero mode and a volume divergence that arises from a momentum integral \cite{Troost:2010ud}. As a consequence, only those contributions that one can isolate from the volume integral have a direct state interpretation in the Ramond sector. These are the pole contributions to the elliptic genus. In the cigar weighted index, there is no single such contribution that survives the $\tau_2 \rightarrow \infty$ limit. Thus, the state interpretation of the result is schematic at best. Still, we can identify the left-moving quantum numbers of the states that contribute to the weighted index. Let us recall a few formulas for the R-charge and conformal dimension of left-moving states on the cigar:
 \begin{align}
 h &= - \frac{j (j-1)}{{k}} + \frac{(m+n_f)^2}{{k}} + \frac{n_f^2}{2}  
\nonumber \\
Q &= -\frac{2(m+n_f)}{{k}} + n_f = -\frac{n+ {k} w}{{k}} + n_f
 \end{align}
 where $j$ is an $sl(2,\mathbb{R})$ spin, $m$ is a left-moving $U(1)$ quantum number and $n_f$ is the fermion number. The quantum numbers $(n,w)$ are a momentum and winding on the cigar. The left-moving quantum numbers that match the charges of the left-moving contributions to the weighted index are:
\begin{align}
(j,n,w,n_f) =\begin{cases} 
(\, \frac{1}{2} \,  , \, 0 \, , \, 0 \, , \,  -\frac{1}{2}) \quad &\text{for}\quad Q = -\frac12\\
(\frac{{k}+1}{2},0,-1, - \frac{1}{2})\quad &\text{for}\quad Q = \frac12~.
\end{cases}
\end{align}
We can split our cigar elliptic genus calculation to exhibit one or the other as a pole contribution, but the split has no invariant meaning -- except in the presence of a non-zero $\beta_2$ chemical potential. 
There is a further aspect of these models that we wish to stress. If we flow left Ramond sector states with charges between $-1/2$ and $+1/2$ to the left NS sector, than these become chiral primaries. These chiral primaries have minimal R-charge equal to $\frac{1}{{k}}$. On the other hand, the NS sector ground state with quantum numbers 
\begin{align}
(j;n,w;n_f) &= (\frac{1}{2};0,0;0) \, ,
\end{align}
has zero R-charge and is not a chiral primary. When flowed back to the Ramond sector, it is an excited state.

\subsection{The Weighted Index of the Orbifold }

For the orbifolded theory, we again begin with the form of the path integral best suited for the calculation of the index:
\begin{align}
\chi^{\mathbb{Z}_P}(\tau, \alpha, \beta )&=
\sqrt{\frac{N\tau_2}{D}}
\sum_{n,w}
 \int_{0}^1 ds_i \frac{\theta_1(\tau,-s_1 \tau-s_2+\alpha)}{\theta_1(\tau,-s_1 \tau-s_2)}
 e^{2\pi i P n s_2} ~ q^{n(Ps_1+w)}
\nonumber\\
&\hspace{2cm} 
e^{2\pi i \alpha(\frac{w}{P} + \frac{D}{N} nP)} ~ e^{-2\pi i \beta P n}  ~(q\bar q)^{-\frac{DP^2}{4N}\big(n-\frac{N}{DP}(\frac{w}{P} + s_1-\beta_2)\big)^2}~.
\end{align}
In our analysis we set $\beta_2=0$. In the limit $\tau_2\rightarrow\infty$, the only contributions  arise from the zero angular momentum sector and from the integration region near $w+s_1 P=0$. There are $P+1$ contributing regions. In order to treat these in a uniform manner, we define the variable $r =s_1 + w/P$ near each solution and substitute for $s_1$. In these regions, we approximate:  
\begin{align}
I^{\mathbb{Z}_P}:= \lim_{\tau_2 \rightarrow \infty} \chi^{\mathbb{Z}_P} &=
\lim_{\tau_2 \rightarrow \infty} \sqrt{\frac{N \tau_2}{D} }  \sum_w \int_{0}^1 ds_2 
\int_{-\epsilon}^{\epsilon} dr ~
 \frac{\theta_1((-\frac{w}{P}+r)\tau + s_2 - \alpha ) }{\theta_1((-\frac{w}{P}+r)\tau + s_2)}
e^{2\pi i \alpha \frac{w}{P} }~
    e^{-\frac{N}{D} \pi \tau_2\, r^2 } \, . \nonumber 
\end{align}
We distinguish between positive and negative $r$ and write:
\begin{multline}
I^{\mathbb{Z}_P} =
\lim_{\tau_2 \rightarrow \infty} \sqrt{\frac{N\tau_2}{D} } \sum_w \int_{0}^1 ds_2 
\int_{0}^{\epsilon} dr e^{-\frac{N}{D} \pi \tau_2\, r^2 } \,  e^{\frac{2\pi iw}{P} \alpha  } \nonumber \\
\times \left( \frac{\theta_1((-\frac{w}{P}+r)\tau + s_2 -\alpha ) }{\theta_11((-\frac{w}{P}+r)\tau + s_2 )}
+ \frac{\theta_1((-\frac{w}{P}-r)\tau + s_2 - \alpha) }
{\theta_11((-\frac{w}{P}-r)\tau + s_2 )}
   \right)~.
\end{multline}
The quantum number $w$ can be either zero or negative. We first consider the cases that $w=0$ and $w=-P$ and find:
\begin{align}
I^{\mathbb{Z}_P}_{0} &= \lim_{\tau_2 \rightarrow \infty} \sqrt{\frac{N\tau_2}{D} } \int_{0}^1 ds_2 
\int_{0}^{\epsilon} dr e^{-\pi \tau_2 \frac{r^2N }{D } } 
\left( \frac{\sin \pi (r\tau + s_2- \alpha ) }{\sin \pi(r\tau + s_2 )}
+ \frac{\sin \pi (-r\tau + s_2 - \alpha ) }{\sin \pi(-r\tau + s_2 )}
   \right)
   \nonumber \\
   &= \lim_{\tau_2 \rightarrow \infty} \sqrt{\frac{N\tau_2}{D} } 
\int_{0}^{\epsilon} dr e^{-\pi \tau_2 \frac{r^2N }{D } } 2 \cos \pi \alpha =   \cos \pi \alpha \, .
\end{align}
This is the same calculation as was done for the coset. For the orbifold theory,  there are more contributions to the index. They arise from the additional winding sectors $-w=\ell=1,2,\dots,(P-1)$. These twisted sectors contribute:
\begin{align}
I^{\mathbb{Z}_P}_{\text{tw}} &=
\lim_{\tau_2 \rightarrow \infty} \sqrt{\frac{N\tau_2 }{D} }   \sum_{l=1}^{P-1} \int_{0}^1 ds_2 
\int_{-\epsilon}^{\epsilon} dr 
 \frac{\sin \pi ((\frac{\ell}{P}+r)\tau + s_2 - \alpha) }{\sin \pi( (\frac{\ell}{P}+r)\tau + s_2)}
   e^{-\pi \tau_2 \frac{r^2N }{D} }  \, e^{-\frac{2\pi i}{P}\ell\alpha  }
   \nonumber \\
   &= \lim_{\tau_2 \rightarrow \infty} \sqrt{\frac{N\tau_2}{D} }   \sum_{l=1}^{P-1} 
\int_{-\epsilon}^{\epsilon} dr 
 e^{\pi i \alpha}
   e^{-\pi \tau_2 \frac{r^2N }{D} }  \, e^{-\frac{2\pi i}{P}\ell\alpha  }
   \nonumber \\
   &=    \sum_{l=1}^{P-1}
 e^{2\pi i (\frac{1}{2}-\frac{\ell}{P}) \alpha} \, .
\end{align}
The weighted index for the $\mathbb{Z}_P$ orbifold is the sum of all contributions and we obtain
\begin{equation}
I^{\mathbb{Z}_P} = \frac{1}{2} z^{- \frac{1}{2}} + z^{- \frac{1}{2}+\frac{1}{P}}
+ \dots+ z^{\frac12 - \frac{1}{P}}
+ \frac{1}{2} z^{ \frac{1}{2}} \, .
\label{WeightedOrbifoldIndex}
\end{equation}
The weighted orbifold index receives contributions from the strictly normalizable  left Ramond ground states with charges equal to 
\begin{equation}
-\frac12 + \frac{1}{P},-\frac12 + \frac{2}{P},\dots, -\frac12 + \frac{P-1}{P} \, . \label{RList}
\end{equation} The discrete states that touch the continuum at charges $-\frac12$ and $\frac12$ each contribute with a factor of one half to the weighted index. As before, their net contribution to the supersymmetric index is one. The total Witten index equals $P$.  

Note that all the charges lie within an interval of size one, independent of the central charge $c$ of the model.
This is in rather sharp contrast to ${\cal N}=2$ minimal models or sigma-models with compact K\"ahler manifold targets where the charges have a width of $c/3$. For non-compact models, the central charge $c$ can be large, but the width of the charge interval remains one.

In the calculation of the elliptic genus, the right-moving Ramond sector cancellation of the fermion zero mode and the volume divergence -- the cancellation of two $\theta_1 (\bar{\tau})$ functions, one arising in the numerator from the Ramond sector and one in the denominator from the non-compact target space volume -- was essential to obtain a well-defined and finite result \cite{Troost:2010ud}. This cancellation does not take place in the NS sector where the trace would be plagued by a volume divergence. An analysis of NS sector chiral primaries is nevertheless useful in that it provides an operator ring. Moreover, the 
(left) operator ring maps (left) Ramond sector ground states to other  Ramond sector ground states as we will see momentarily.

\subsection{A Landau-Ginzburg Perspective on the Ground States}
We provide further intuition for our results using a Landau-Ginzburg or Liouville perspective. 
Consider a ${\cal N}=2$ Liouville theory at central charge
\begin{equation}
c =3 + \frac{6}{{k}}~,
\end{equation}
and introduce the parameter
\begin{equation}
b = \sqrt{\frac{{k}}{2}} \, .
\end{equation}
We parameterize the Liouville potential in the exponential form:
\begin{equation}
W (\Phi) = \mu e^{b \Phi} = \mu e^{\sqrt{\frac{{k}}{2}} \Phi} \, \label{Superpotential}
\end{equation}
depending on a scalar ${\cal N}=(2,2)$ chiral multiplet $\Phi$. 
The slope of the linear dilaton equals 
\begin{equation}
Q'=\sqrt{\frac{2}{k}} \, .
\end{equation}
The minimal period for the canonically normalized compact scalar component $\theta$ in $\Phi=\rho+i \theta$ equals $R_{L,min}=Q'$. This model is FZZ dual \cite{Kazakov:2000pm, Hikida:2008pe} to the cigar coset discussed in Section \ref{EllipticGenus}, i.e. the coset theory that is regular near the tip and which has the maximal asymptotic radius. 

In the Liouville description, the radius can be taken to be any integer multiple of $Q'$. For a multiple $R_{L}=PQ'$, we have that the Witten index of the model equals $P$ by a simple path integral argument \cite{Girardello:1990sh}. With such a radius, the $\theta$ momentum becomes quantized in units of $1/(P Q')$. All these reasonings are valid for a generic  positive level ${k}$.  The Witten index agrees with the $z=1$ value of the  weighted orbifold index (\ref{WeightedOrbifoldIndex}) of the cigar. Indeed, the $\mathbb{Z}_P$ orbifold reducing the radius of the cigar by a factor of $P$ is T-dual to the Liouville theory with $P$ times the minimal radius. 

The $(1/2,1/2)$ chiral-chiral operator equal to the superpotential $W$ (\ref{Superpotential}) always exists in the Liouville theory and has the quantum numbers:
\begin{align}
(j,m,\bar{m},h,Q,n_f,n_L) &= (\frac{k}{2},\frac{k}{2},\frac{k}{2},\frac12,1,0,1) \, .
\end{align}
For the minimal radius model, it has the smallest Liouville angular momentum $n_L=1$. It is T-dual or FZZ dual to the winding one condensate in the cigar model. When we orbifold by the group $\mathbb{Z}_P$, there are other chiral-chiral operators in the Liouville theory that have a fractional angular momentum quantized in units of $1/P$.  The operators can be described as:
\begin{equation}
O_l = W^{ \frac{l}{P}} \, ,
\end{equation}
where $l=1,\dots,P-1$. These operators form an operator ring which we can divide out by the ideal generated by $W'$.\footnote{See  \cite{Li:2018rcl} for a detailed discussion in the case of integer level.} In the exponential parameterization, this leaves us with the operators with exponents of the superpotential smaller than one.  Again, this reasoning is valid at generic level ${k}$.
When we  act with these operators on the (almost) normalizable left-moving Ramond ground state of minimal R-charge, we will generate the list of states with the charges listed in (\ref{RList}) that contribute to the elliptic genus.

\section{Conclusions}
\label{Conclusions}
In this paper we further explored the zoo of elliptic genera of non-compact conformal field theories, the mathematical objects to which they give rise, as well as their interpretations. Firstly, we observed that the three parameter elliptic genera in the literature could be extended to any value of the central charge of the ${\cal N}=2$ superconformal cigar theory between three and infinity. The elliptic genus is generically modular covariant and elliptic in the angular fugacity. Because of the generically irrational ratios of  R-charges, the elliptic genus is a non-holomorphic function of three variables that generalizes Jacobi forms. 

Secondly, we  enlarged the space of flavored elliptic genera of orbifold theories, thus generalizing the orbifolded genera in \cite{Sugawara:2011vg}. We argued that the natural starting point of our analysis is the elliptic genus of the cigar coset. The asymptotic radius of this model is set by the level $k$, while the geometry is regular and smooth at the tip. All orbifolded genera are obtained as angular orbifolds of this model. To compute the elliptic genera, it was instrumental to have the three-parameter, flavored elliptic genus of the cigar at hand. At a geometric level, what the $\mathbb{Z}_P$ orbifold does is to introduce an orbifold singularity near the tip, while asymptotically it shrinks the radius of the circle direction of the cigar by a factor of $1/P$. There is an FZZ dual description in terms of ${\cal N}=2$ Liouville theory, in which the cigar at maximal radius is mapped to the Liouville theory at minimal radius. In this Liouville description, the $\mathbb{Z}_P$ orbifold enlarges the  radius of the asymptotic circle to $P$ times the minimal radius. It is a well known result \cite{Girardello:1990sh} that the Witten index of this model is equal to $P$. We confirmed this by taking the appropriate limit of the path integral expression for the $\mathbb{Z}_P$-orbifolded genus. In fact we performed a detailed analysis of the Witten index weighted by the R-charge of the Ramond ground states, and made manifest the subtleties associated to the continuum of the spectrum touching the discrete states. 

Starting from the path integral expression for the orbifolded elliptic genus, we also exhibited the holomorphic part and the remainder using standard techniques. The most interesting  aspect of this analysis is the fact that  the remainder term  we obtain is closely related to the modular invariant partition function for the asymptotic rational compact boson. In particular, for a  generic orbifold, the remainder term is a non-diagonal combination of the left and right sectors. This is a generalization of the existing results for elliptic genera of non-compact models, giving rise to new completed mock Jacobi forms with modular and elliptic properties that depend intricately on the positive integers parameterizing this class of models\footnote{See \cite{Bhand:2023rhm} for another interesting generalization of mock modular forms \cite{Zwegers}, to meromorphic Jacobi forms of rational index.}.  

One can imagine applications of the conformal field theories we studied, as string theory backgrounds. Fractional level cigar models arise for instance in backgrounds T-dual to highly curved NS5-brane backgrounds. Moreover, 
the orbifold models that we studied have applications as non-compact Gepner models in string theory. An original class of backgrounds is the class in which one combines two cigar models at central charges that sum to an integer, for instance. It is natural to study orbifolds thereof, just as it is in traditional Gepner models. 

Independent of further applications though, 
the careful analysis of the flavored orbifolded cigar models illustrates a more general known theme that deserves  further exploration. To what extent can simple physical models and their observables, and in particular the elliptic genera of non-compact superconformal field theories, give rise to interesting classes of mathematical objects worthy of study in their own right? Apart from their applications in string theory, anomalies and black hole physics, these interesting  objects may stand the test of time and find entirely different applications.  

\appendix 

\section{Extended Characters}
\label{extendedcharacters}

The twisted ${\cal N}=2$ superconformal Ramond character for a  representation built on a Ramond sector ground state of R-charge $Q$ is:
\begin{equation}
\text{ch}_d^{\tilde{R}} (Q;\tau,\alpha) 
=  
z^{Q
} ~\frac{z^{1/2}}{1-z} ~\frac{-i\theta_1(\tau, \alpha)}{\eta(\tau)^3}
\, .
\end{equation}
For a central charge $c=3+6D/N$ theory, the extended characters at level $N$ are obtained by a weighted sum over spectral flowed copies of the ground state. They are labelled by the level $N$, the R-charge of the Ramond ground state and the spectral flow intercept $a$:
\begin{align}
    \text{Ch}_d^{\tilde{R}} (N, v,a;\tau,\alpha,\beta) &= \sum_{n \in a+ N\mathbb{Z}} q^{\frac{c}{6}n^2} ~z^{\frac{c}{3}n}
    ~y^{-n}
    ~\text{ch}_d^{\tilde{R}} (Q=\frac{v}{N}-\frac12; \tau, \alpha + n\tau)
\nonumber \\
& = -\frac{i\theta_1(\tau,\alpha)}{\eta(\tau)^3} \sum_{n \in \mathbb{Z}} q^{ND (n+\frac{a}{N})^2} ~z^{2D (n+\frac{a}{N})}~
 (zq^{Nn+a})^{\frac{v}{N}} \frac{y^{-Nn-a} }{1-z q^{Nn+a}} 
  ~.
\end{align}
We used the variables $z=e^{2 \pi i \alpha}$ and $y=e^{2 \pi i \beta}$.
The extended characters are a function of the modular parameter $\tau$, the R-charge fugacity $\alpha$ as well as a fugacity $\beta$ that keeps track of the spectral flow copy. It will play the role of the angular momentum fugacity in the bulk of the paper. For the  elliptic genera of $\mathbb{Z}_P$ orbifolds, we need more refined extended characters at level $NP$ which we define as:
\begin{align}
    \text{Ch}_d^{\tilde{R}} (NP, v,a;\tau,\alpha,\beta) &= \sum_{n \in a+ P N\mathbb{Z}} q^{\frac{c}{6}n^2} ~z^{\frac{c}{3}n}
    ~y^{-n}
    ~\text{ch}(Q=\frac{v}{P N}-\frac12, \tau, \alpha + n\tau)
    \\
    &=-\frac{i\theta_1(\tau,\alpha)}{\eta(\tau)^3}  \sum_{n \in \mathbb{Z}} q^{DN P^2 (n+\frac{a}{PN})^2} ~z^{2DP (n+ \frac{a}{NP})} (zq^{NP n + a})^{\frac{v}{PN}} \frac{y^{-NPn-a}}{1-z q^{NPn+a}}     ~. \nonumber
\end{align}
These characters are labelled by an integer $v$ which will take values in $\{0,1,\dots,NP \}$ and which correspond to Ramond sector ground states of charge between $-1/2$ and $+1/2$. 

\bibliographystyle{JHEP}

\end{document}